# Antenna enhanced graphene THz emitter and detector

Jiayue Tong[1*], Martin Muthee[2*], Shao-Yu Chen[1], Sigfrid K. Yngvesson[2] and Jun Yan[1†]

[1]Department of Physics, University of Massachusetts, Amherst, Massachusetts 01003, USA

[2]Department of Electrical and Computer Engineering, University of Massachusetts, Amherst, Massachusetts 01003, USA

[†]Corresponding Author: Jun Yan.    Tel:   (413)545-0853    Fax:   (413)545-1691
E-mail: yan@physics.umass.edu

* These authors contributed equally to this work.

**Abstract**

Recent intense electrical and optical studies of graphene have pushed the material to the forefront of optoelectronic research. Of particular interest is the few terahertz (THz) frequency regime where efficient light sources and highly sensitive detectors are very challenging to make. Here we present THz sources and detectors made with graphene field effect transistors (GFETs) enhanced by a double-patch antenna and an on-chip silicon lens. We report the first experimental observation of 1-3 THz radiation from graphene, as well as four orders of magnitude performance improvements in a GFET thermoelectric detector operating at ~2 THz. The quantitative analysis of the emitting power and its unusual charge density dependence indicate significant non-thermal contribution from the GFET. The polarization resolved detection measurements with different illumination geometries allow for detailed and quantitative analysis of various factors that contribute to the overall detector performance. Our experimental results represent a significant advance towards practically useful graphene THz devices.

**Manuscript text**

The gapless electronic structure of graphene[1] is a unique property that has drawn significant attention from both basic sciences and practical applications[2]. In particular, it enables broadband interaction of photons with the two dimensional (2D) atomic layer from the far infrared up to the ultraviolet[3]. This has led to various optoelectronic devices operating with photons in the visible[4–7], near infra-red[8–11], mid infra-red[12–16] and far infra-red[17–24]. Applications of graphene field effect transistors (GFET) in the few terahertz (THz) frequency range are particularly appealing since it's one of the least developed regimes lying in the gap between efficient manipulation with electronics and photonics[25–27]. Here we perform combined THz emission-detection measurements using devices made with monolayer graphene. Our results represent the first study of THz emission from graphene, as well as significant improvements in GFET thermoelectric THz detectors.

A common bottleneck in graphene photonic and optoelectronic devices is the limited light-matter interaction, because of the 2D crystal's sub-nanometer thickness. This has led to the 'greybody' radiation[28,29] and limited responsivity in graphene photon detectors[30]. Various routes have been employed to improve the performance of graphene optoelectronic devices, such as, photon gating effects[31–33], introducing electron trapping centers[34], placing the device inside an optical cavity[35,36], on a waveguide[37–40] or photonic crystal[41,42], coupling with an antenna[22,23], or gold nanostructures[43,44], utilizing plasmon resonances[16,18,45], etc.. In this work we make use of a double-patch antenna and an on-chip silicon lens to enhance the coupling between THz light and graphene. Significant performance improvements are observed for emission from electrically-biased graphene transistors, and for detection with un-biased graphene devices, at the few THz frequency range that is notoriously difficult to work with. For the emitter, we observe a radiated power that is significantly larger than the anticipated thermal radiation, suggesting additional radiation channels at our disposal for devising efficient graphene THz sources. For the detector, we achieve four orders of magnitude sensitivity improvements, which, in conjunction with its high speed[19,20], makes the GFET a strong competitor to other contemporary THz sensors.

The antenna is designed to have a size of 45×31 μm$^2$ as shown in Fig.1 (a). The graphene piece (black line) is positioned at the central gap in-between the two metal patches made by palladium or gold. The 45 μm long slot at the location of the graphene channel is chosen such that it can accommodate well our target frequency at about 2THz (corresponding to a wavelength of 150 μm in free space and 45 μm in silicon). We use HFSS (High Frequency Structure Simulator) to simulate the performance of the antenna coupled GFET circuitry (see Methods). The calculated impedance spectrum in Fig.1(b) indicates an optimal operation frequency of 2.1THz. The electric field distribution at the optimal frequency is shown in Fig.1(c) where it is seen that the oscillating fields are peaked at the center and the two edges of the slot, corresponding to a full wavelength of the THz light.

The graphene-antenna circuitry is fabricated on a high resistivity silicon chip with a 300nm-thick silicon oxide dielectric. The chip is subsequently glued back-to-back on an elliptical silicon lens using purified bee wax. The double patch antenna that connects the graphene at THz also serves as the source and drain electrodes for the graphene transistor. For emission measurements (Fig.1d) an electrical current *I* is applied to heat up the device. Radiation in the target THz range is coupled out by the antenna through the silicon lens and is detected either directly or spectroscopically. Conversely, for detection measurements as illustrated in Fig.1e, THz radiation from an external laser source is coupled into the antenna through the silicon lens and dissipated by the graphene. We detect the voltage difference $V$ between the source and drain electrodes generated during this dissipation process to quantify the detector sensitivity as normalized by the total incident power.

Figure 2 a,b show the top optical image of device D1 and the side schematics of the graphene THz emitter (fabrication details in Methods). We use a custom-built Fourier-transform THz spectrometer (FTS; see Methods) shown in Fig.2c to analyze the spectral distribution of the THz radiation from the electrically heated graphene. The measured radiation spectrum in Fig.2d strongly peaks around 1.9 THz. The two side dips are caused by ambient water absorption (grey overlapping bands). The overall lineshape of the response curve agrees reasonably with the simulation (dashed curve, calculated from the simulated impedance in Fig.1b), while the experimental peak position is shifted towards a lower frequency. We investigated the gate voltage/charge density dependence of the spectra. As shown in Fig.2d, when more holes are introduced into the device (see Fig.2e for gate dependence of resistance), some spectral weight is transferred from the higher energy shoulder (blue curve) to the lower energy shoulder (pink curve).

We further measured the gate voltage $V_G$ dependence of total THz radiated power (Fig.2e black curve; 1V corresponds to $7.2 \times 10^{10} \text{cm}^{-2}$ of carrier density change) directly with the silicon bolometer, bypassing the FTS paths in Fig.2c. Interestingly, the radiated power of the device under a constant bias voltage displays an obvious peak, while the electrical impedance increases monotonically as $V_G$ is increased. This suggests that there exists a THz radiative channel in GFET that is peaked at a certain charge density. We speculate that this could be related to plasmonic effects in the graphene that have attracted significant recent attention[12–14,17,18,50–55].

To quantitatively analyze the THz emission from the coupled graphene-antenna circuit, we measured device D2 both its temperature and its integrated THz emission power (see Methods). As shown in the Raman spectra of Fig.3a, the Stokes G and 2D phonon bands soften upon heating with an electric current, and the anti-Stokes G band (inset) emerges from the flat background at room temperature, consistent with previous experiments[28]. The integrated intensity ratio of the Stokes $I_S$ and anti-Stokes $I_{aS}$ bands allows us to determine the temperature of the optical phonon $T_{op}$: $\frac{I_{aS}}{I_S} = C\exp\left(-\frac{\hbar\omega_G}{k_B T_{op}}\right)$ where $\hbar\omega_G$=196meV is G band phonon energy, $k_B$ is the Boltzmann constant, and $C \approx$ 0.86 is a calibration constant determined empirically (see Methods). From data in Fig.3a we found $T_{op} \approx$440K at an electrical heating power of 27mW. Since the G band is strongly coupled to the electron degrees of freedom[46], we assume that $T_{op}$ is equal to the electron temperature as has been verified by previous experiments[28].

The integrated THz emission power is measured by placing the device in front of a calibrated Golay cell nominally sensitive to photons in the 0.15-20THz frequency range. To ensure that only radiation in the target 1-3THz is collected, we performed large THz range FTS characterization and verified that there is no detectable radiation from 3 up to 20THz. Photons at even higher frequencies are blocked by a Zitex G106 sintered Teflon filter[47]. The integrated 1-3 THz radiated power scales linearly with DC heating as shown in Fig.3b. The radiated power reaches 2.1nW at a heating power of 27mW, i.e. an electron temperature of 440K.

The quantitative determination of the temperature and the corresponding radiated power in Fig.3 allows us to estimate the efficiency of the device. We first hypothesize that the radiation is of thermal origin. Using an equivalent circuit, we model the graphene as a resistive load connected to an antenna radiating thermal noise[48]. From the antenna theory[48] and since the 1-3THz photon energies are small compared with $k_BT$, the ideal radiated power is given by $k_BT \cdot B$, where the bandwidth $B$ is about 1THz estimated from the measured spectra in Fig.2d. Our measurement system detects the THz power difference between the GFET and the room temperature background (see Methods), giving an anticipated power of $k_B \cdot (440K-293K) \cdot 1THz = 2nW$ in the absence of any loss. The silicon lens is known to have a reflection loss of about 30%[49]. The impedance mismatch loss is given by $\left(\frac{Z_0-Z_d}{Z_0+Z_d}\right)^2$, where $Z_0$ is the antenna impedance which is about 50Ω (see the peak value in Fig.1b), and $Z_d$ is the graphene electrical impedance measured to be 1800Ω for device D2. This results in a loss of 88%. We neglect the loss due to radiation into the sample side (experimentally we did observe some radiated power when the sample side is facing the Golay cell; see Methods), as well as the absorption loss due to ambient moisture, the silicon substrate and the silicon lens at room temperature. Based on these considerations, the anticipated upper bound of the THz thermal radiated power is 0.2nW. The surprisingly 10-fold larger THz power that we observe in Fig.3b, together with the unusual gate voltage/charge density dependence in Fig.2e, suggest that the radiation mechanism is not limited to the Nyquist thermal noise and highlights the role of graphene played in the device.

It is interesting to compare our results with earlier experiments on electrically-biased graphene emitters in the visible[28] and near infrared[29] range. In these GFET thermal emission investigations, it was found that in the absence of antenna coupling, the emissivity is about 2.3%[56–58] and the graphene photon emitter can be described as a 'greybody'. If we extrapolate the greybody radiation in Refs. [28] and [29] to the THz range in our measurements, the 10μm² flake of graphene typically used in our experiments would only produce 1.8pW THz radiation at 440K, three orders of magnitude smaller than the 2.1nW observed in our experiment. In this context the antenna coupled graphene device investigated here is a very efficient radiation source.

The graphene-antenna-silicon lens system is also employed for THz detection. In these experiments the direction of the THz radiation is reversed (Fig.1e vs. 1d) and graphene becomes a THz absorber instead of a radiator. The absorption heats up electrons and diffusion of hot carriers establishes in the device a temperature gradient $\nabla T$. This creates a thermoelectric field $E = -S\nabla T$ where $S$ is the Seebeck coefficient. The voltage that we measure in Fig.1e is given by a line integration of the electric field along the device $V = \int_{So}^{Dr} E \cdot dl = -\int_{So}^{Dr} S\nabla T \cdot dl$ where *So* and *Dr* represent the source and drain electrodes which serve as heat sinks.

From the above equation for the thermoelectric voltage it is apparent that the detected *V* would approach zero if *E* near the source and drain electrodes were to have similar magnitude and opposite signs. Similar cancellation effects have been experimentally observed by previous graphene photo-detector studies[5,10]. To efficiently break the source-drain mirror symmetry, we devised a device geometry shown in Fig.4a: the source electrode is palladium and makes a 2D contact with the surface of the graphene, while the drain is chromium/gold and makes a 1D contact with the graphene edge[59], using a thin atomic film of boron nitride to protect the graphene flake (see Methods). This makes use of the work function difference between different metals, as well as the significant contact resistance difference between surface and edge contacts[59]. We found this device geometry so far gives us the best detection performance (see Fig.4b for optical image of the device D3 made this way).

To take advantage of the measured spectral response information in Fig.2d, a matching 1.9 THz gas laser output is employed for the far infrared detection measurements (see Methods). Two different illumination geometries are investigated: front illumination (Fig.4c upper inset, coupling through the silicon lens) and back illumination (Fig.4c lower inset, directly onto the graphene and the antenna). For each measurement we recorded the detailed response dependence on the direction of the THz electric field with respect to the antenna. The detected thermoelectric voltage is normalized to the total incident power, which is measured simultaneously with a beam splitter and a commercial detector to correct any power fluctuations during the experiment.

The responsivity for front illumination (red curve in Fig.4c) is more than two orders of magnitude larger than that for back illumination (black curve), as calculated in the F-B ratio in Fig.4d. In the same illumination geometry, the sensitivity is markedly improved when the electric field is perpendicular to the slot, in agreement with HFSS simulation that finds little field strength parallel to the slot. The responsivity can be further enhanced by changing the gate voltage/carrier density. As shown in Fig.4f, the THz detector is strongly carrier density tunable and peaks at $V_G \approx 40V$. The 'S' shaped response curve, in contrast to the resistance in Fig.4e that peaks at the charge neutral Dirac point, is reminiscent of the Seebeck coefficient studies of graphene[60], in agreement with the thermoelectric origin interpretation of the response.

Our maximum detection responsivity of 4.9 V/W in Fig.4f represents an encouraging development in GFET THz detectors. Despite recent progress in the sub-THz range[22,23], GFET detector responsivity is still fairly low at photon frequencies of a few THz. A recent 2.5 THz detection study finds a responsivity of 0.48mV/W (referenced to the 17mW total incident power, or 10V/W referenced to the 0.75μW THz power on the graphene[19]). This is comparable to the lowest responsivity of 0.35mV/W we observe in Fig.4c. The antenna and the silicon lens are seen to improve the response to 1.1V/W at zero gate voltage, and $V_G$ tuning in Fig.4f further increases it to 4.9V/W, representing a 4 orders of magnitude improvement over Ref.[19].

To assess the ultimate performance of the GFET thermoelectric detector, we perform a DC (direct current) rectification measurement to find the electric heating responsivity of the same detector D3. A DC bias voltage is used to heat the sample and the thermoelectric current is detected by comparing the magnitude of the electric currents for the two opposite directions of bias voltage (see Methods). This provides an independent and intrinsic measurement of the thermoelectric response of the sample. As shown in Fig.4g, it also displays a similar 'S' shape, but the magnitude is more than two orders of magnitude larger than the THz responsivity observed in Fig.4f. This implies that despite the improvements we have observed, there is still significant room for further optimization of the graphene THz detector. Such optimization should also lead to improvements for the THz emitter presented in Figs. 2&3.

The signal-to-noise ratio of the detector is characterized by the noise equivalent power (NEP) of the device. In Fig. 5a we plot the measured gate dependence of the Johnson noise of the sample (black dots), which shows reasonable agreement with, albeit slightly larger than, the theoretical estimations of the room temperature Johnson-Nyquist noise floor $\sqrt{4k_B TR}$ with $R$ being the resistance measured in Fig.4e. The NEP of the detector is calculated from the ratio of the noise voltage in Fig.5a and the optical voltage responsivity in Fig.4f. As shown in Fig.5b, the NEP reaches a minimum level of 1.7nW·Hz$^{-1/2}$ at the peak responsivity.

In conclusion, we have studied an integrated silicon lens-antenna-graphene transistor system both as emitters and as detectors in the technologically important few THz spectral range. In both applications significant performance improvements are observed due to the enhanced coupling. Our first demonstration of graphene THz emission opens up new opportunities to further investigate plasmon enhanced graphene THz radiation and to create highly efficient and coherent THz light sources. The versatile measurements of the thermoelectric detector delineate in detail how each factor in the design impacts the overall response of the device in a quantitative manner, and pave way for engineering and optimizing each contributing factor leading to commercial graphene THz detectors.

**Methods**

**Simulation.**

The antenna simulation was performed using ANSYS HFSS software, a 3D full-wave electromagnetic field solver. The antenna and part of the leads were designed in the simulation to match the actual fabricated device. The graphene sheet was modelled as a lumped port with $Z_o$ equal to the DC resistance at the peak THz power under a gate voltage. The location of the lumped port feed matched the graphene sheet location in the actual device as determined from optical microscope inspection.

**Device Fabrication**.

Graphene emitter: Monolayer graphene is mechanically exfoliated on high resistivity (>1 KΩ·cm) silicon chips with a 300nm oxidized dielectric layer. The layer thickness is identified by optical contrast and verified by Raman spectroscopy. The source and drain antenna contacts are patterned with electron beam (e-beam) lithography followed by either e-beam evaporation of titanium/gold (5/100nm) or sputtering of palladium (30nm). Data in Fig.2 are from device D1 and data in Fig.3 are from device D2.

Graphene detector: All the detector data in Figs. 4 and 5 are taken from device D3 (optical image in Fig.4b). The graphene-boron nitride heterostructure is made by transferring mechanically exfoliated boron nitride on top of a monolayer graphene piece using poly-propylene carbonate following Ref.[59]. A PMMA etching mask is defined with e-beam lithography to selectively expose one side of the graphene-boron nitride atomic stack for reactive-ion etching using $CHF_3$ and $O_2$ gases. The exposed graphene edge is subsequently contacted with chromium/gold metallization[59]. For the other side of the device, we make a conventional 2D surface contact by sputtering palladium.

**Raman Spectroscopy**.

The room temperature ambient Raman measurement was performed with a Horiba T64000 spectrometer equipped with a liquid nitrogen cooled CCD. The spectrum of electrically biased graphene was excited by an Argon laser (488nm) and collected with a 40× objective lens (NA=0.6). The optical system is calibrated with the silicon phonon at $520cm^{-1}$.

**THz Emission Measurements.**

The spectrum of the THz emission: the sample is heated by an HP 8116A function generator. THz radiation is analyzed using a customized Fourier-transform THz spectrometer (FTS; see Fig. 2c): the collimated light is split by a mylar beam splitter into two arms with the beam paths fixed for one arm and slowly varied for the other. The interference of photons from the two arms is detected by a liquid helium cooled silicon bolometer. Fourier transform of the recorded interferogram to the frequency domain gives the THz spectrum shown in Fig.2d. While Fig.2d only shows data below 3.5THz, we did measure up to 20THz and observed no radiation above 3THz.

Integrated THz power measurement: Heating of the sample is achieved with the same HP 8116A function generator. Current in the sample is modulated at 5Hz and the power we detect is the difference between the heated graphene and the room temperature background. A Zitex G106 filter is placed in-between the device and a calibrated Golay cell to efficiently block photons above 12THz[47]. All the emission measurements in Figs.2&3 are performed with front illumination, i.e. from the silicon lens side. We also checked the back illumination geometry, i.e. with the sample side facing the Golay cell, and did observe some, albeit much weaker and more divergent, THz radiation.

**Graphene THz Detection.**

The responsivity of the graphene THz detector is measured with the 158μm (1.9THz) laser line of diflouromethane ($CH_2F_2$) gas pumped by a $CO_2$ laser at 10.6μm. The laser beam is chopped at 20Hz with a mechanical chopper and the signal is detected with a lock-in amplifier. We used two linear polarizers followed by a beam splitter in front of the sample to perform polarization dependent detection. The linear polarizers enable us to change the power and the polarization of THz electric field with respect to the antenna direction. We measure simultaneously responses from the graphene detector and a commercial pyroelectric detector using the two beams after the beam splitter. This enables us to normalize the device response to the incident laser power in-situ when the light polarization is tuned.

**Electrical measurements.**

The device resistance (Fig.2e, Fig.4e) and Johnson noise (Fig.5a) are measured with a lock-in amplifier (10nA current excitation) and a HP 35665A Dynamic Signal Analyzer (zero bias), respectively. The DC gate voltage is varied by a Keithley 2400 multimeter.

The DC heating responsivity in Fig.4g is measured with a Stanford SIM 928 isolated voltage source to minimize the voltage noise. We apply DC voltages with the same magnitude and opposite directions across the device and measure the corresponding current in the two cases. When the bias voltage switches sign, the bias current reverses direction while the thermoelectric current remains in the same direction. The rectification current (current difference under forward and reverse bias) multiplied by the sample resistance gives the thermoelectric voltage, which is subsequently normalized by the electric heating power, allowing us to extract the thermoelectric voltage responsivity in Fig.4g.

**Contributions**

S.K.Y. and J.Y. conceived the experiment. J.T. fabricated the graphene THz emitters and detectors. S.-Y.C. assisted in fabricating graphene and boron nitride samples. J.T. and M.M. performed the emission and detection measurements. M.M. performed the HFSS simulations. J.T. performed the Raman and electrical measurements. All authors participated in analyzing the data and writing the paper.

**Acknowledgements**

This work is supported by the University of Massachusetts Amherst, the National Science Foundation Center for Hierarchical Manufacturing (CMMI-1025020) and the Materials Research Science and Engineering Center on Polymers (DMR-0820506).

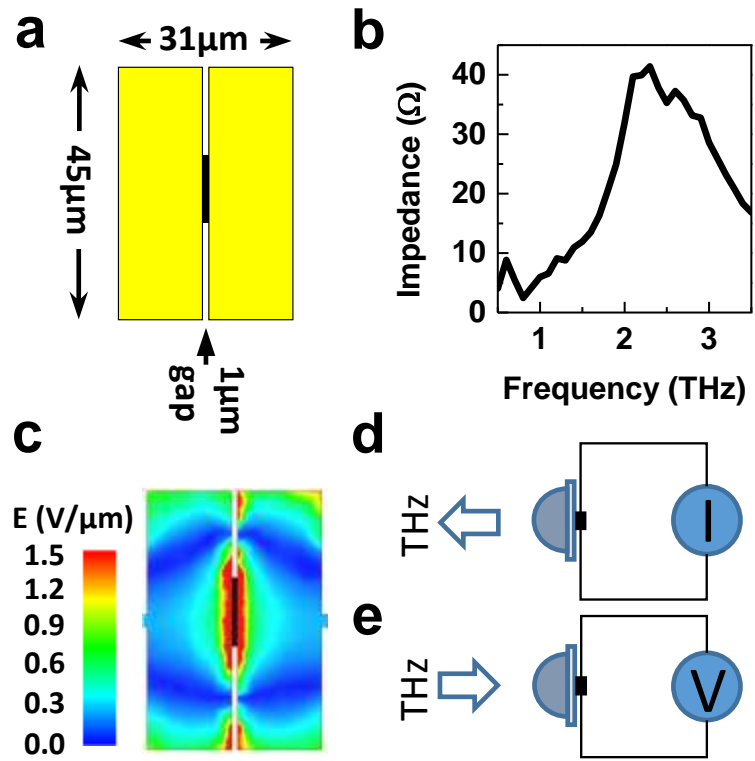

**Figure 1**. (**a**) Design of the THz antenna coupled graphene device. (**b**) HFSS simulated frequency dependence of impedance. (**c**) Calculated distribution of THz electric field at 2.1THz. (**d, e**) Schematic setup of emission (**d**) and detection (**e**) experiments.

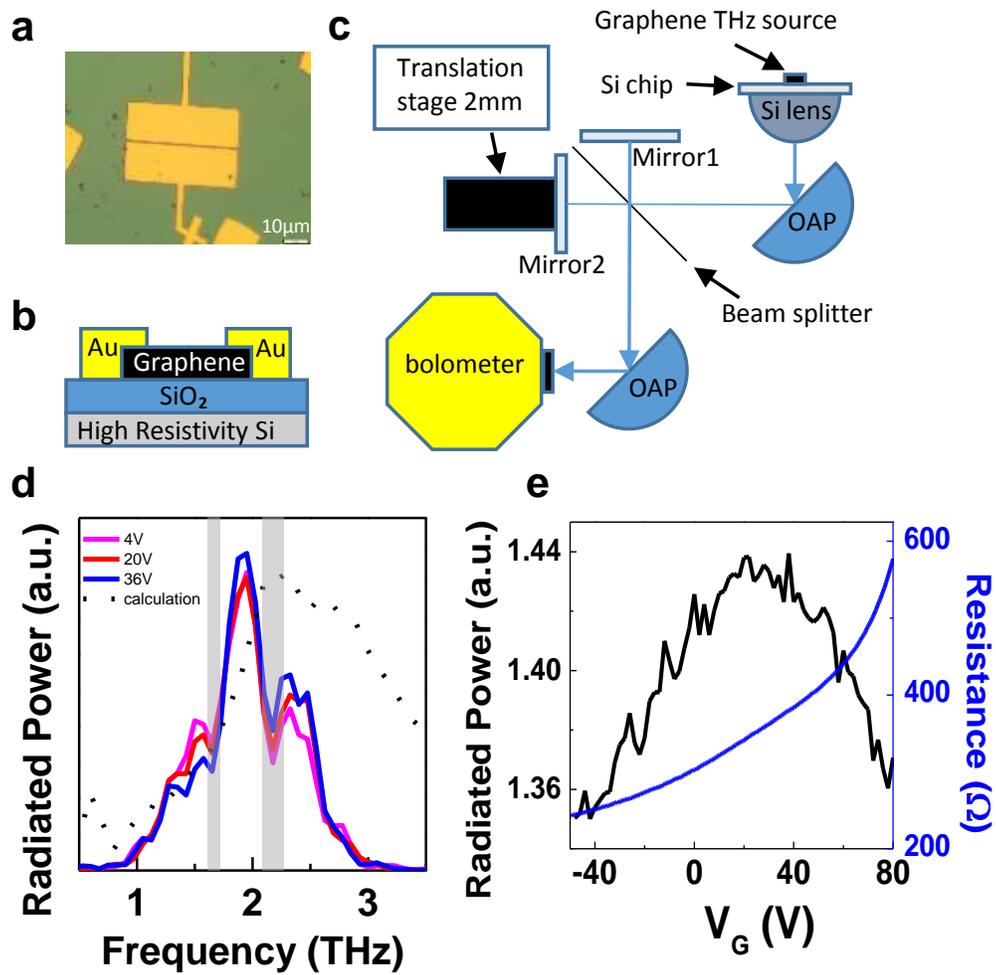

**Figure 2.** (**a**) Optical image of the GFET emitter. (**b**) Schematic side view of GFET emitter. (**c**) Fourier-transform THz interferometer set-up. (**d**) Measured (solid lines) THz radiation from graphene at different gate voltages. The dashed line is the transmission spectrum calculated from HFSS simulation in Fig.1b. The two grey bands indicate water absorption lines. (**e**) Comparison of $V_G$ dependent GFET resistance (blue) and THz radiated power (black).

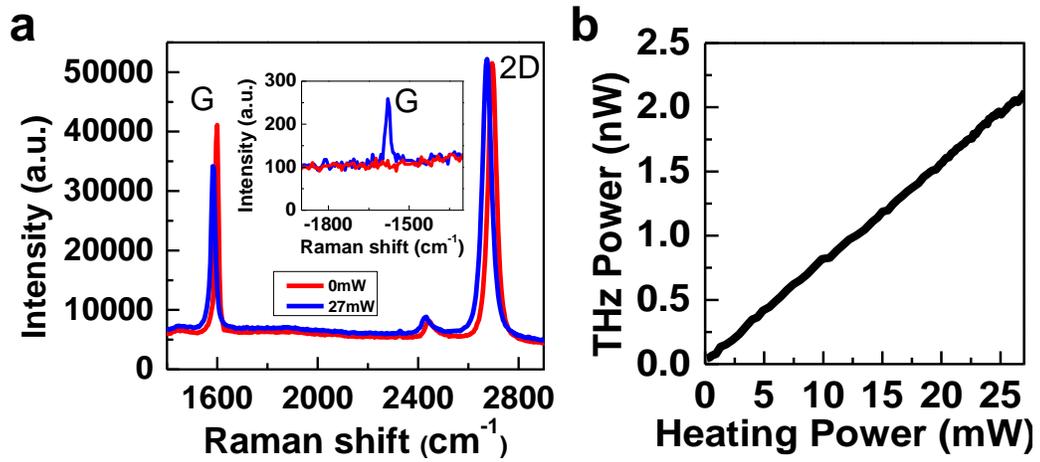

**Figure 3.** (**a**) Stokes (main panel) and anti-Stokes (inset) Raman spectra of a GFET in the absence (red) and presence (blue) of bias current. (**b**) Dependence of THz radiated power on DC electrical heating.

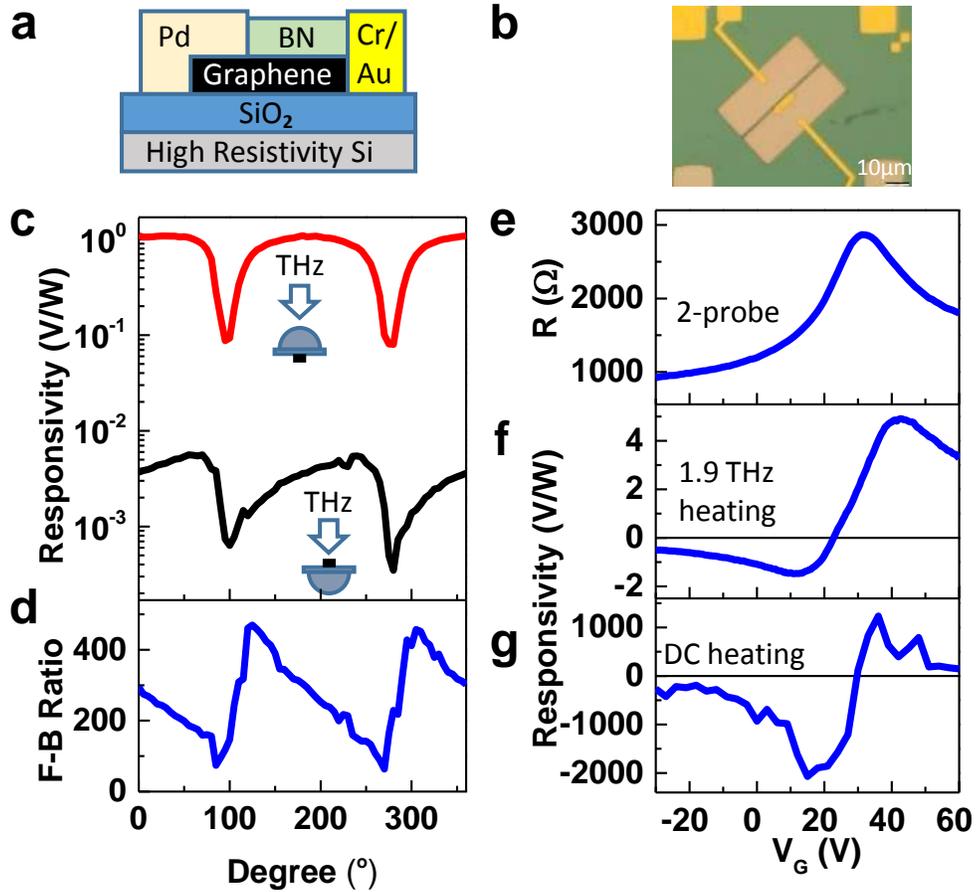

**Figure 4.** (**a**) Schematic side view of the THz detector design. (**b**) Optical image of the detector. (**c**) Polarization dependence of the responsivity at zero gate voltage under front illumination (red) and back illumination (black). Insets illustrate the illumination geometry. (**d**) Ratio of the responsivities for the two illumination geometries as a function of the polarization angle. (**e**) Resistance of the detector as a function of the gate voltage. (**f**) Detector responsivity to 1.9THz radiation as a function of the gate voltage. (**g**) Detector responsivity to Joule heating as a function of the gate voltage.

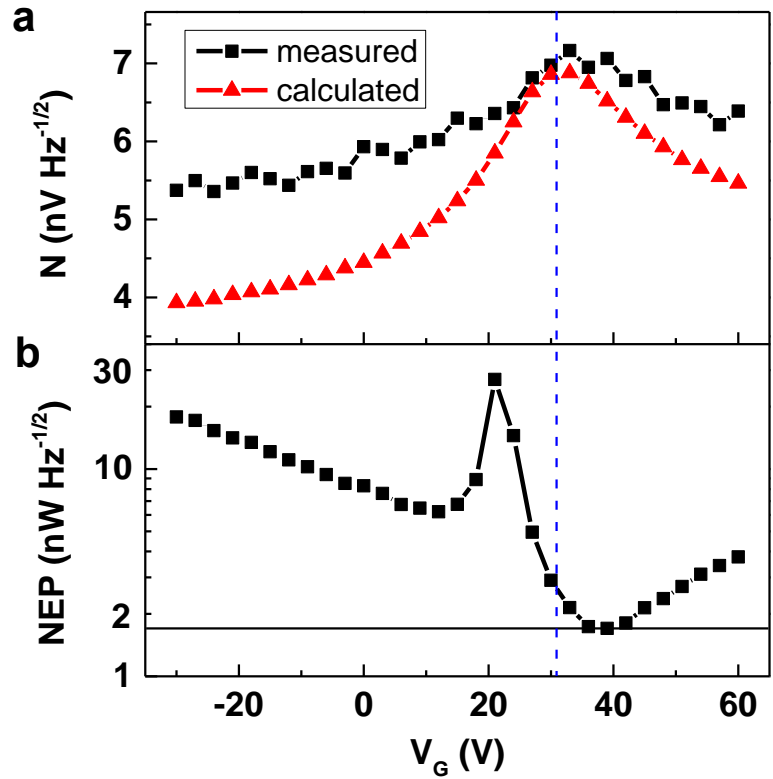

**Figure 5.** (**a**) Measured (black dots) and calculated (red dots) Johnson-Nyquist noise voltage of the THz detector. (**b**) The noise-equivalent power (NEP) of the graphene detector.